\documentclass[journal]{IEEEtran}
\usepackage{cite}

\ifCLASSINFOpdf
   \usepackage[pdftex]{graphicx}
   \graphicspath{{fig/}}
   \DeclareGraphicsExtensions{.png}
\else
  \usepackage{graphicx}
\fi
\usepackage{amsmath}
\usepackage{amsfonts}
\newcommand{\mat}[1]{\mathbf{#1}}

\usepackage{algorithmic}
\ifCLASSOPTIONcompsoc
 \usepackage[caption=false,font=normalsize,labelfont=sf,textfont=sf]{subfig}
\else
 \usepackage[caption=false,font=footnotesize]{subfig}
\fi

\usepackage{xcolor}

\begin{document}
%
\title{Non-negative Matrix Factorization using Partial Prior Knowledge for Radiation Dosimetry}

%
%
%

\author{Boby~Lessard,
        Frédéric~Marcotte,
        Arthur Lalonde,
        François~Therriault-Proulx,
        Simon~Lambert-Girard,
        Luc~Beaulieu,
        and~Louis~Archambault
\thanks{This work did not involve human subjects or animals in its research.}
\thanks{B. Lessard, is with the Département de physique, génie physique et optique, et Centre de recherche sur le cancer, Université Laval, as well as with the CHU de Québec – Université Laval et CRCHU de Québec, and with Medscint inc., Québec, Canada.}
\thanks{F. Marcotte, L. Beaulieu and L. Archambault are with the Département de physique, génie physique et optique, et Centre de recherche sur le cancer, Université Laval, as well as with the CHU de Québec – Université Laval et CRCHU de Québec, Québec, Canada.}
\thanks{F. Therriault-Proulx and S. Lambert-Girard are with Medscint inc., Québec, Canada.}
\thanks{A. Lalonde is with the Département de Radio-Oncologie, CHUM, with the Centre de recherche du CHUM, as well as with the Département de Physique, Université de Montréal, Montréal, Canada.}
}

\maketitle

\begin{abstract}
Hyperspectral unmixing aims at decomposing a given signal into its spectral signatures and its associated fractional abundances. To improve the accuracy of this decomposition, algorithms have included different assumptions depending on the application. The goal of this study is to develop a new unmixing algorithm that can be applied for the calibration of multi-point scintillation dosimeters used in the field of radiation therapy. This new algorithm is based on a non-negative matrix factorization. It incorporates a partial prior knowledge on both the abundances and the endmembers of a given signal. It is shown herein that, following a precise calibration routine, it is possible to use partial prior information about the fractional abundances, as well as on the endmembers, in order to perform a simplified yet precise calibration of these dosimeters. Validation and characterization of this algorithm is made using both simulations and experiments. The experimental validation shows an improvement in accuracy compared to previous algorithms with a mean spectral angle distance (SAD) on the estimated endmembers of 0.0766, leading to an average error of $(0.25 \pm 0.73)$\% on dose measurements.
\end{abstract}


%
\IEEEpeerreviewmaketitle

\section{Introduction}
\label{sect:intro}
%
%
%
%
%

\IEEEPARstart{H}{yperspectral} unmixing is a concept used in fields such as geoscience, computer vision, biophotonics and astrophysics \cite{keshava_spectral_2002, qu_hyperspectral_2018, brownstein_biophotonic_2007, shen_spatial-spectral_2012}.  
The aim of hyperspectral unmixing is to identify the individual components of a mixed signal. 
This can be achieved by exploiting the distinct spectral signature of each component.
A hyperspectral sensor, such as a spectrometer or a hyperspectral camera, will measure light over multiple narrow wavelength bands to produce a single spectrum containing signal from multiple sources. The challenge of hyperspectral unmixing is to decompose that spectrum into two parts: the spectral signature of every component (referred to as \textit{endmembers}), and their related fractional \textit{abundances} \cite{keshava_spectral_2002}.

Hyperspectral unmixing has been applied to the field of medical physics for precise measurements of ionizing radiation doses with single-point and multi-point scintillation dosimeters (mPSDs) in the context of radiation therapy \cite{archambault_mathematical_2012, therriault-proulx_development_2012}. 
Radiation therapy is a cancer treatment modality that uses beams of ionizing radiation to irradiate cancerous cells, while sparing healthy tissues as much as possible. 
Radiation doses are typically delivered by a medical linear accelerator (linac). 
Radiation treatment delivery must be carefully monitored to ensure patient safety and treatment efficiency. The requirements in precision and accuracy for different aspects of radiation therapy are codified in reports such as the Task Group 142 (TG-142) of the American Association of Physicists in Medicine (AAPM) \cite{klein_task_2009}.
In this context, plastic scintillation detectors (PSDs) possess many advantages that are well suited for modern radiation treatments, such as water-equivalence and real-time dose reading \cite{beddar_water-equivalent_1992, beddar_water-equivalent_1992-1}, which are assets for dose measurements in radiotherapy.
Multi-point plastic scintillation detectors have been proposed to expand the capabilities of traditional PSDs \cite{therriault-proulx_development_2012}. 
These detectors consist of multiple scintillating elements each emitting a different emission spectrum, coupled to a single optical fiber. 
These scintillating elements can be arranged in a waterfall configuration, meaning that all scintillators are on the same optical line, or they can be on separate heads that ultimately connect to a single optical line.
A raw measurement performed with such dosimeter consists of measuring the superposed spectra produced by the scintillators and optical fiber during its exposition to ionizing radiation \cite{therriault-proulx_nature_2013}. 
Because the light produced by a scintillator is proportional to the radiation dose it receives, determining the abundance of a given scintillator element is equivalent to evaluating the dose received by that element.
Determining the contribution of each component requires hyperspectral unmixing
\cite{archambault_mathematical_2012, therriault-proulx_development_2012}, because at least 3 spectral components (scintillation, fluorescence and Cherenkov light) must be considered if only one scintillator is present in the dosimeter. In the context of a mPSD, like the one developed by Linares-Rosales \textit{et al.} \cite{linares_rosales_optimization_2019} with all scintillators on the same optical line for \textit{in-vivo} applications, the use of hyperspectral unmixing becomes even more essential because more than one scintillation spectrum is present within the overall signal. 
However, determination of these spectral signatures is not an easy task to perform \cite{therriault-proulx_development_2012} and better unmixing algorithms are necessary to improve the accuracy of dose determination. 

Many unmixing algorithms have been developed which may or may not require the presence of pure measurements (i.e. measurements composed of only one endmember) \cite{tang_nonnegative_2012}. 
Among unmixing algorithms that do not require the presence of pure measurements, we find the non-negative matrix factorization (NMF) \cite{pauca_nonnegative_2006} that expresses the measurement matrix by the product of two matrices: the endmember matrix and the abundance matrix. 
Starting with an initial estimation of the endmember and abundance matrices, a minimization algorithm, such as a gradient descent, is used to simultaneously determine the two matrices to be estimated.
The natural non-negativity constraint of the NMF makes it interesting to use for hyperspectral unmixing because the spectral signatures and abundances of a spectral measurement should always be positive.
However, this factorization does not have a unique solution and the algorithm can be trapped in local minima due to a non-convex objective function \cite{tong_nonnegative-matrix-factorization-based_2016}. 
Consequently, the solution depends strongly on the initial estimations of both matrices. 
To help reach the global minimum of the objective function, other algorithms such as the non-negative double singular value decomposition (NNDSVD) \cite{boutsidis_svd_2008} can be applied to get a good initialization of the NMF algorithm.

To address the problem of non-uniqueness of the solution and considering that some endmembers are known perfectly while others are unknown, Tang \textit{et al.} \cite{tang_nonnegative_2012} developed the NMF-UPK algorithm that includes partial \textit{a priori} information on endmembers. 
When some of the endmembers are accurately known, this method helps retrieving the unknown endmembers with more accuracy than if no prior information was provided.
However, this method is not well suited for cases where there can be discrepancies between the known endmembers and the actual endmembers, because the known endmembers are fixed and can't evolve during the iterations of the algorithm.
Sigurdsson \textit{et al.} \cite{sigurdsson_endmember_2014} proposed an alternative method allowing the known endmembers to deviate from their initial shape using soft regularization in order to retrieve the actual endmembers that could differ slightly from their known values. 
Tong \textit{et al.} \cite{tong_nonnegative-matrix-factorization-based_2016} later applied the same logic, but in a more straightforward way that allowed extending this method to other NMF-based unmixing algorithms. 
Wang \textit{et al.} \cite{wang_multiple_2020} developed the MCG-NMF algorithm that extracts prior information about the endmembers and abundances of the hyperspectral images by performing multiple clustering analysis on the image. 
This information can then be used as partial prior on the endmembers and abundances.

In medical physics applications, endmembers determination can be done using a series of measurements, each with a different vector of fractional abundances, in order to build a calibration dataset. 
However, because each measurement taken with a mPSD is independent from the others, there is no spatial information that can be used to help guide the algorithm. This is different from most geoscience applications where the spatial structure of a hyperspectral image can be used to guide the estimation of endmembers. 
Furthermore, the size of a typical dataset produced by a mPSD typically contains between 10 to 100 data points, which is much smaller than datasets used in geoscience that can contain thousands of pixels. 
Finally, mPSD measurements are not often dominated by only one endmember \cite{therriault-proulx_development_2012}, which can make the unmixing process difficult to perform. 
However, partial prior knowledge can be known in advance on both the endmembers and the abundances, by performing a certain calibration routine. 

In this context, we are proposing a new NMF algorithm, the Non-Negative Matrix Factorization - Partial Endmember-Abundance Knowledge (NMF-PEAK), that considers prior information on both the endmembers and the abundances with an adaptive uncertainty coefficient.
The proposed algorithm intended for the calibration of \textcolor{black}{mPSDs} must be capable of retrieving the abundances to a high level of accuracy on subsequent measurements, because clinical practice usually \textcolor{black}{requires measurements} with 1\% or better accuracy.  
The remainder of the paper is divided as follows: Sec. \ref{sect:LMM-NMF} reviews the theory of the linear mixing model and the basic NMF algorithm, and describes the application of these models to the field of radiation therapy. 
The proposed approach for the new unmixing algorithm is described in Sec. \ref{sect:approach}, while the validation of the algorithm on simulated and experimental data is described in Sec. \ref{sect:validation}. 
Finally, we conclude the results of our work in Sect. \ref{sect:conclusion}.

\section{Linear Mixing Model and NMF Algorithm}
\label{sect:LMM-NMF}

\subsection{Theory of the Linear Mixing Model}

The linear mixing model (LMM) applied to hyperspectral unmixing assumes the $m^{th}$
measured spectrum $\mat{y}_{m,[L\times 1]}$, having $L$ wavelength channels, is a
linear superposition of $K$ endmembers with some stochastic additive noise
$\mat{n}_{m,[L\times 1]}$. Each endmember, $\mat{r}_{k,[L\times 1]}$, is weighted
by the abundance $x_{k, m}$:

\begin{equation}
    \mat{y}_m = \sum_{k=1}^{K} x_{k, m} \mat{r}_{k} + \mat{n}_m 
    = \mat{R} \mat{x}_m + \mat{n}_m\,,\label{eq:y}
\end{equation}

\noindent
where $x_{k,m}$ is the abundance of endmember $k$ for measurement $m$ and
$\mat{R}_{[L\times K]}$ is the endmember matrix. Abundances are expressed as
relative fractions of each endmember in a given measured spectrum, $\mat{y}_m$;
as such, all $y_{\ell,m}$ must be non-negative and the sum of $y_{\ell, m}$ over $L$
must equal to one. The additive noise on measurements can generally be assumed to
be Gaussian. Thus, the noise vector can be expressed as:

\begin{equation}
    \textcolor{black}{\mat{n}_m \sim \mathcal{N}(\mathbf{0},\boldsymbol{\Sigma}_n)\,,} \label{eq:noise}
\end{equation}
where $\boldsymbol{\Sigma}_n$ is the covariant matrix. If the noise is
independent identically distributed (iid), it can be characterized by a
standard deviation $\sigma_n$ such that $\boldsymbol{\Sigma}_n =
\sigma_n^2\mathbf{I}$.

For $M$ measurements, as in the case of a hyperspectral image, eq.~(\ref{eq:y})
can be rewritten as:

\begin{equation}
    \mat{Y} = \mat{R}\mat{X} + \mat{N}\,,
\end{equation}

\noindent
where $\mat{Y}_{[L\times M]}$ is the measurement matrix, each column
representing a single spectral measurement, $\mat{R}_{[L \times K]}$ is the
endmember matrix, $\mat{X}_{[K \times M]}$ is the abundance matrix and
$\mat{N}_{[L\times M]}$ is the noise matrix.

The goal of NMF is thus to determine $\mat{X}$ and $\mat{R}$ from a set of
hyperspectral measurement, \textcolor{black}{$\mat{Y}$.}

\subsection{Application to Scintillation Dosimetry}
When a mPSD is irradiated by a beam of ionizing radiation, each scintillator
component emits light proportionally to the dose deposited within \textcolor{black}{its}
respective volume \cite{beddar_water-equivalent_1992, beddar_water-equivalent_1992-1}. The optical fiber used for guiding the
scintillation light also emits light when irradiated:
fluorescence~\cite{therriault-proulx_nature_2013} and Cherenkov
radiation~\cite{boer_optical_1993}. These emissions are typically not
proportional to the radiation dose and are generally an undesired, parasitic,
contribution to the total signal~\cite{beddar_water-equivalent_1992}. 

By assuming that the LMM is a valid approximation, the signal from a mPSD can be
expressed as the superposition of the light produced by each of the scintillator
elements, the fluorescence light and the Cherenkov radiation. Each measurement
acquired with a mPSD can be compared to a pixel in a hyperspectral image, and
the dose for one measurement can be determined if the fractional abundance of
each scintillating element is known~\cite{archambault_mathematical_2012}. 

The calibration of a mPSD consists in finding the endmembers (\emph{i.e.}
$\mat{R}$ in eq.~(\ref{eq:y})). Once this is known, the abundance can be
determined for any measurement $\mat{y}_m$ by multiplying it by the pseudo-inverse
$\mat{R}^+$. Because of the proportional relationship between the radiation dose
and the amount of light emitted by a scintillator, the abundance of each
scintillator element can be directly linked to the dose received by that
element if the total emitted light is known. 
The calibration has to be performed the first time a mPSD is used and
on a regular basis \textcolor{black}{afterwards}, because the endmembers may change over time due to
aging and damages caused by radiation. 

One way to obtain the endmembers and abundances is to perform an extensive
calibration of the dosimeter where the spectrum of each endmember is measured
accurately. This is practically challenging, because isolating each endmember
when irradiating the mPSD is a complex task. Furthermore, such calibration is a
long and tedious process that requires a specialized procedure which is not
practical in every clinical setting. To facilitate this calibration for
day-to-day clinical applications of \textcolor{black}{mPSDs}, we propose to retrieve the endmembers
using a NMF-based algorithm that uses prior information combined with a fast and
simple experimental procedure. 

\begin{figure}[!t]
\centering
\includegraphics[width=3.25in]{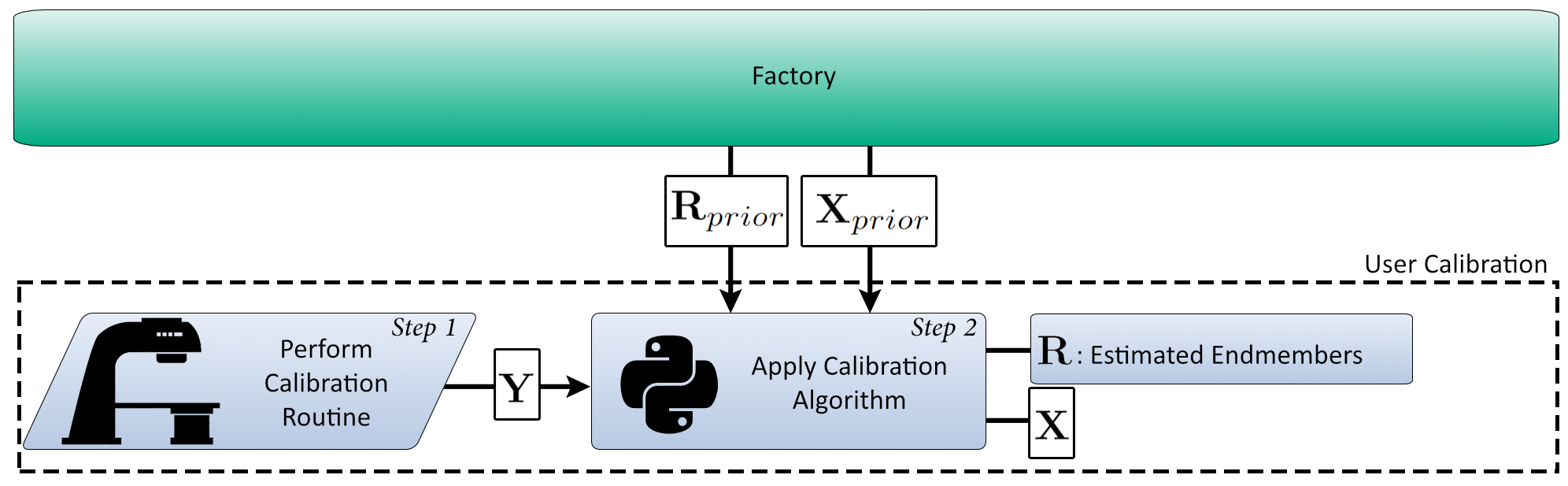}
\caption{Experimental process applied by the user to perform the calibration of the mPSD using a calibration algorithm.}
\label{fig:process_simple}
\end{figure}

This is a two-step process. 
\emph{Step~1}: 
A simple calibration routine consisting of a few irradiations in conditions that are easy to reproduce is performed by the end user in a clinical setting. \emph{Step~2}: The NMF-based algorithm is applied, using $\mat{R}_\mathrm{prior}$ and
$\mat{X}_\mathrm{prior}$ as prior information to guide the NMF-based algorithm. The priors on $\mat{R}$ are obtained, for example, by following the method described in appendix \ref{app:long_calib}. It could also be estimated from numerical simulations for example. The priors on $\mat{X}$ are obtained by performing the same simple calibration routine as the one to be performed by the end user, and using $\mat{R}_\mathrm{prior}$ to compute the abundances. The determination of $\mat{R}_\mathrm{prior}$ and $\mat{X}_\mathrm{prior}$ could be performed only once at the
factory by the manufacturer or in a non-clinical setting for example.
This process is illustrated on Figure \ref{fig:process_simple}.

\section{Proposed Approach for Spectral Unmixing}
\label{sect:approach}

\subsection{Proposed Objective Function for the New NMF-based Algorithm}

Here, we derive the objective function of the algorithm from a Bayesian
perspective, following an approach similar to
\cite{bioucas-dias_hyperspectral_2012, dobigeon_joint_2009,
moussaoui_separation_2006, arngren_bayesian_2009}. According to Bayes' theorem,
the spectral unmixing problem can be expressed as

\begin{equation}
    P(\mathbf{R},\mathbf{X}|\mathbf{Y}) = \frac{P(\mathbf{Y}|\mathbf{R},\mathbf{X})P(\mathbf{R})P(\mathbf{X})}{P(\mathbf{Y})}\,,
\label{eq:bayes}
\end{equation}
where $P(\mathbf{R},\mathbf{X}|\mathbf{Y})$ is the joint posterior probability
density, $P(\mathbf{Y}|\mathbf{R},\mathbf{X})$ is the likelihood function
depending on the measurements $\mathbf{Y}$ given the endmembers $\mathbf R$ and
the abundances $\mathbf{X}$ and $P(\mathbf{R})$ and $P(\mathbf{X})$ are the
priors of the endmembers and abundances respectively.

While it is possible to define prior distributions that satisfy the LMM
unmixing constraints (positive abundances, sum to one) in order to perform
Bayesian inference~\cite{dobigeon_joint_2009}, such approaches require complex
sampling strategies~\cite{theys-2009}. Here, the Bayesian framework is used only
to derive the objective function for maximum a posteriori
estimation~\cite{parra_unmixing_2000}.

Given that the noise on a measurement is assumed Gaussian as described by
eq.~(\ref{eq:noise}), the likelihood of observing $\mat{y}_m$ given $\mat{x}_m$
and $\mat{R}$ should be the same Gaussian
distribution~\cite{dobigeon_joint_2009}. Thus, the likelihood of observing the
set of $M$ measurements $\mat{Y}$, assuming that the noise is independent of the
measurement sequence, is: 

\begin{align}
    P(\mat{Y}|\mat{R},\mat{X}) &= \prod_{m=1}^M \frac{\exp\left[-\frac{1}{2\sigma^2}\|\mat{y}_m - \mat{Rx}_m\|^2\right]}{\left(2\pi\sigma^2\right)^{L/2}} \nonumber \\
    &= \textcolor{black}{\left(2\pi\sigma^2\right)^{-LM/2}
        \exp\left[-\frac{\|\mat{Y} - \mat{RX}\|^2}{2\sigma^2}\right]\,,}
    \label{eq:prior_noise}
\end{align}
where $\| \cdot \|^2$ is the Frobenius norm.

To define the distributions $P(\mathbf{R})$ and $P(\mathbf{X})$, we propose
multivariate Gaussian functions centered respectively at $\mathbf{R}_\mathrm{prior}$
and $\mathbf{X}_\mathrm{prior}$. This implies that both the endmembers and the
abundances of the calibration routine should be normally distributed around the
reference values.

\begin{align}
    P(\mat{R}) &= \prod_{k=1}^{K} \frac{
        \exp{\left[-\frac{1}{2\sigma_k^2} \left\| (\mat{r}_k - \mat{r}_{k, \mathrm{prior} }) \right\|^2\right]}}
        {(2\pi\sigma_k^2)^{L/2}} \nonumber \\
        &=  \frac{
        \exp{\left[-\frac{1}{2} \left\| (\mat{R} - \mat{R}_\mathrm{prior})\mat{\Sigma}_\mat{R}^{-1/2} \right\|^2\right]}}
        {(2\pi)^{LK/2}\det\left(\mat{\Sigma}_\mat{R}\right)^{L/2}}\,,\\
    &\nonumber \\
    P(\mat{X}) &= \prod_{m=1}^{M} \frac{
        \exp{\left[-\frac{1}{2\sigma_m^2} \left\| (\mat{x}_m - \mat{x}_{m, \mathrm{prior} }) \right\|^2\right]}}
        {(2\pi\sigma_m^2)^{K/2}} \nonumber \\
        &=  \frac{
        \exp{\left[-\frac{1}{2} \left\| (\mat{X} - \mat{X}_\mathrm{prior})\mat{\Sigma}_\mat{X}^{-1/2} \right\|^2\right]}}
        {(2\pi)^{KM/2}\det\left(\mat{\Sigma}_\mat{X}\right)^{K/2}}\,.
\end{align}

Since the $K$ endmembers are assumed to be independent, $\mat{\Sigma}_\mat{R}$
is a diagonal matrix. The elements of that matrix are $K$ variance parameters,
$\sigma^2_k$, of the Gaussian distribution used to model $P(\mat{R})$. The larger
$\sigma_k$ is, the wider the probability distribution is for endmember $k$.
Thus, the quantity $1/\sigma_k$ can be used as a parameter
representing the confidence in the prior value of endmember $k$. We define the
matrix $\mat{A}$ as 

\begin{equation}
    \mathbf{A}_{[K \times K]}^2 = \sigma^2 \mathbf{\Sigma_R}^{-1} = 
    \left(
    \begin{array}{ccc}
    \alpha_{1}^2 & \cdots & 0 \\
    \vdots & \ddots & \vdots \\
    0 & \cdots & \alpha_{K}^2
    \end{array}
    \right)\,,
\end{equation}
where the $\alpha_{k}$, with $k \in \{1, 2, ..., K\}$, are referred to as \emph{trust coefficients on prior endmembers}.

Similarly, we define
\begin{equation}
    \mathbf{B}_{[M \times M]}^2 = \sigma^2 \mathbf{\Sigma_X}^{-1} = 
    \left(
    \begin{array}{ccc}
    \beta_{1}^2 & \cdots & 0 \\
    \vdots & \ddots & \vdots \\
    0 & \cdots & \beta_{M}^2
    \end{array}
    \right)\,,
\end{equation}
where the $\beta_{m}^2$, with $m \in \{1, 2, ..., M\}$, can be interpreted as a confidence parameter on the prior
measurements from the calibration routine. Because these measurements are used
to determine $\mat{X}_\mathrm{prior}$, the $\beta_m$ are henceforth referred to
as \textit{trust coefficients on prior abundances}.

Having defined the probability density of eq.~(\ref{eq:bayes}), we can estimate
the abundances and endmembers with the following cost function:

\begin{equation}
    \textcolor{black}{\mat{X}, \mat{R} = \underset{\mat{X},\mat{R}}{\mathrm{arg}\,\mathrm{max}}\,
    P(\mathbf{R},\mathbf{X}|\mathbf{Y})\,.} \label{eq:argmax}
\end{equation}

Taking the logarithm of~(\ref{eq:argmax}) and neglecting the constant terms
yields the following function to minimize:

\begin{equation}
\begin{aligned}
    F(\mathbf{R},\mathbf{X}) = \frac{1}{2} &\parallel \mathbf{Y} - \mathbf{RX} \parallel^2 \\
    + \frac{1}{2} &\parallel (\mathbf{R} - \mathbf{R}_\mathrm{prior})\mathbf{A} \parallel^2 \\
    + \frac{1}{2} &\parallel (\mathbf{X} - \mathbf{X}_\mathrm{prior})\mathbf{B} \parallel^2\,,
\end{aligned}
\label{eq:costfct}
\end{equation}
where the $\alpha_k$
and $\beta_m$ coefficients are relative to the first term, because they included $\sigma^2$ in their definition.

A gradient descent can be used to minimize eq.~\ref{eq:costfct}. Lee and
Seung~\cite{lee_algorithms_nodate} have derived the multiplicative update rule
(MUR), which uses a gradient descent with an adaptive step size chosen so that
the update rules, for both matrices, become multiplicative. This type of update
rule is a good compromise between ease of implementation and rapidity of
convergence~\cite{lee_algorithms_nodate}, and can be implemented for any
$\beta$-divergence objective functions, including the Frobenius
norm~\cite{fevotte_algorithms_2011-1}. The MUR and a hierarchical alternating
least squares algorithm (HALS)~\cite{cichocki_fast_2009} have been compared.

The gradient of the objective function to be minimized can be expressed as
\begin{equation}
    \begin{split}
        \nabla F(\mathbf{R},\mathbf{X}) = \biggl[
        &\frac{\partial F}{\partial \mathbf{R}_{11}}\,,
        ...\, ,
        \frac{\partial F}{\partial \mathbf{R}_{LK}}\,, \\
        &\frac{\partial F}{\partial \mathbf{X}_{11}}\,,
        ...\,,
        \frac{\partial F}{\partial \mathbf{X}_{KM}}
        \biggr]_{[P\times 1]},
    \end{split}
    \label{eq:gradient}
\end{equation}
where $P = LK + KM$. In equation~(\ref{eq:gradient}), $\partial F/\partial
\mathbf{R}_{lk}$ and $\partial F/\partial \mathbf{X}_{km}$ correspond to the
elements of the following matrices, having dimensions $L \times K$ and $K \times
M$ respectively:
\begin{align}
    \frac{\partial F(\mathbf{R},\mathbf{X})}{\partial \mathbf{R}} &= -\mathbf{Y} \mathbf{X}^T + \mathbf{RX} \mathbf{X}^T + \left[\mathbf{R} - \mathbf{R}_\mathrm{prior}\right] \mathbf{A}^2\,,
    \label{eq:grad_R}
    \\
    \frac{\partial F(\mathbf{R},\mathbf{X})}{\partial \mathbf{X}} &= -\mathbf{R}^T \mathbf{Y} + \mathbf{R}^T \mathbf{RX} + \left[\mathbf{X} - \mathbf{X}_\mathrm{prior}\right]\mathbf{B}^2.
    \label{eq:grad_X}
\end{align}

The gradient descent is performed by updating $\mathbf{R}$ and $\mathbf{X}$
using the update rules:
\begin{align}
    \mathbf{R}_{lk}^{(n+1)} &= \mathbf{R}_{lk}^{(n)} - \gamma_{lk} \frac{\partial F(\mathbf{R},\mathbf{X})}{\partial \mathbf{R}_{lk}}\,,
    \label{eq:update_R}
    \\
    \mathbf{X}_{km}^{(n+1)} &= \mathbf{X}_{km}^{(n)} - \gamma_{km} \frac{\partial F(\mathbf{R},\mathbf{X})}{\partial \mathbf{X}_{km}}
    \label{eq:update_X}\,,
\end{align}
where $\gamma_{lk}$ and $\gamma_{km}$ are the step sizes of the gradient
descent. Matrices $\mathbf{R}_\mathrm{prior}$ and $\mathbf{X}_\mathrm{prior}$ do
not have to be fully defined before applying the algorithm. To account for
missing information, we set $\alpha_k^2 = 0$ (or $\beta_m^2 = 0$) for the
corresponding endmember (or measurement abundance) in the hyperparameter matrices
$\mathbf{A}$ and $\mathbf{B}$.

\subsection{Multiplicative Update Rule Algorithm}

The step sizes in equations~(\ref{eq:update_R}) and~(\ref{eq:update_X}) can be
chosen in order to get a MUR~\cite{lee_algorithms_nodate}. In that case, the
following step sizes are proposed:

\begin{align}
    \gamma_{lk} &= \frac{\mat{R}_{lk}^{(n)}}{(\mat{RXX}^T)_{lk}^{(n)} + \alpha_k^2 \mat{R}_{lk}^{(n)}}\,,\label{eq:step_R_mu} \\
    \gamma_{km} &= \frac{\mat{X}_{km}^{(n)}}{(\mat{R}^T\mat{RX})_{km}^{(n)} + \beta_m^2 \mat{X}_{km}^{(n)}}\,.\label{eq:step_X_mu}
\end{align}

The update rules (equations~(\ref{eq:update_R}) and (\ref{eq:update_X})) then
become:

\begin{align}
    \mathbf{R}_{lk}^{(n+1)} &= \mathbf{R}_{lk}^{(n)} \cdot \left(\frac{(\mat{YX}^T)_{lk}^{(n)} + \alpha_k^2(\mat{R}_\mathrm{prior})_{lk}}{(\mat{RXX}^T)_{lk}^{(n)} + \alpha_k^2\mat{R}_{lk}^{(n)}}\right)\,,
    \label{eq:MUR_R}
    \\
    \mathbf{X}_{km}^{(n+1)} &= \mathbf{X}_{km}^{(n)} \cdot \left(\frac{(\mat{R}^T\mat{Y})_{km}^{(n)} + \beta_m^2(\mat{X}_\mathrm{prior})_{km}}{(\mat{R}^T\mat{RX})_{km}^{(n)} + \beta_m^2\mat{X}_{km}^{(n)}}\right)\,.
    \label{eq:MUR_X}
\end{align}

This algorithm is non-increasing, which means that the value of the objective function does not increase from one iteration to the other.
The proof of this non-increasing property is demonstrated in appendix \ref{app:proof_nip_MUR}. However, as outlined by authors such as Berry \textit{et al}.
\cite{berry_algorithms_2007} \textcolor{black}{and} Chih-Jen \cite{chih-jen_lin_convergence_2007},
this non-increasing property does not assure the convergence of the algorithm to
a saddle point. This can also be seen from equations~(\ref{eq:MUR_R}) and
(\ref{eq:MUR_X}), where the algorithm stops to update the values of
$\mat{R}_{lk}$ or $\mat{X}_{km}$ if $\mat{R}_{lk} = 0$ or $\mat{X}_{km} = 0$.

\subsection{Hierarchical Alternating Least Squares \textcolor{black}{Algorithm}}

Another optimization approach for equation~(\ref{eq:costfct}) is to use a
Hierarchical Alternating Least Squares (HALS)~\cite{cichocki_fast_2009}, which
exploits the fact that the objective function is convex for $\mat{R}$ only or
$\mat{X}$ only, but not for both. The algorithm will minimize alternatively the
objective function~(\ref{eq:costfct}) for a constant $\mat{X}$ and for a
constant $\mat{R}$. The update rules for the HALS algorithm are found by
equating the gradient to 0 in equations~(\ref{eq:grad_R}) and~(\ref{eq:grad_X}),
and are given by:

\begin{align}
    \mat{R}_{lk}^{(n+1)} &= \frac{(\mat{Y}^{(k)} \mat{X}^T)_{lk} + \alpha_k^2(\mat{R}_\mathrm{prior})_{lk}}{(\mat{XX}^T)_{kk} + \alpha_k^2}\,, 
    \label{eq:update_HALS_R}
    \\
    \mat{X}_{km}^{(n+1)} &= \frac{(\mat{R}^T \mat{Y}^{(k)})_{km} + \beta_m^2(\mat{X}_\mathrm{prior})_{km}}{(\mat{R}^T\mat{R})_{kk} + \beta_m^2}\,,
    \label{eq:update_HALS_X}
\end{align}
where $\mat{Y}^{(k)}$ is the residue for the $k^{\textnormal{th}}$ endmember, which depends on $\mat{r}_k$ and $\mat{x}_k$, the column vectors representing the $k^{\textnormal{th}}$ column of matrix $\mat{R}$ and the $k^{\textnormal{th}}$ row of matrix $\mat{X}$ respectively:
\begin{equation}
    \textcolor{black}{\mat{Y}^{(k)} = \mat{Y} - \sum_{k' \neq k} \mat{r}_{k'} \mat{x}_{k'}^T \,.}
    \label{eq:residue}
\end{equation}

The update rules of equations~(\ref{eq:update_HALS_R})
and~(\ref{eq:update_HALS_X}) correspond to the gradient descent for
$\mat{R}_{lk}$ and $\mat{X}_{km}$ (equations~(\ref{eq:update_R})
and~(\ref{eq:update_X})), with an adaptive step size given by:

\begin{align}
    \gamma_{lk} &= \frac{1}{(\mat{XX}^T)_{kk} + \alpha_k^2} = \frac{1}{\mathcal{R}_{kk}}\,, \label{eq:step_R_cd} \\
    \gamma_{km} &= \frac{1}{(\mat{R}^T\mat{R})_{kk} + \beta_m^2} = \frac{1}{\mathcal{X}_{m,kk}}. \label{eq:step_X_cd}
\end{align}

\subsection{Implementation of the \textcolor{black}{Algorithm}}

The proposed NMF-based algorithm has the advantage to be easy to implement from
a pre-existing NMF-based algorithm. It was implemented in \textit{Python}, based
on the NMF algorithm from the Scikit-Learn library~\cite{scikit-learn}. The
initialization of the algorithm is performed using the method NNDSVDA which
performs a non-negative double singular value decomposition from the given
matrix $\mat{Y}$ and thereafter replaces the null values by the average value of
$\mat{Y}$. However, one advantage of incorporating the priors on endmembers and
abundances is that the \textcolor{black}{algorithm does} not depend on the initialization method as
much as it does with the basic NMF algorithm. Moreover, the initialization can
also be performed using the prior knowledge on endmembers or abundances if they
are sufficiently known. Once the initialization is done, the gradient descent is
performed alternatively on $\mat{R}$ and $\mat{X}$ using either the step sizes
defined in equations~(\ref{eq:step_R_mu}) and~(\ref{eq:step_X_mu}) (MUR
algorithm) or the ones defined in equations~(\ref{eq:step_R_cd})
and~(\ref{eq:step_X_cd}) (HALS algorithm), depending on the choice of the user. A projection step is also applied to
ensure the non-negativity of matrices $\mat{R}$ and $\mat{X}$. We also note that
the HALS algorithm is implemented in \textit{Cython}, compared to the MUR
algorithm. For both update algorithms, the convergence criterion is reached when
a specific metric becomes smaller than $10^{-10}$ times the initial metric. For
the MUR algorithm, the metric corresponds to the difference between the value of
the objective function between 10 consecutive iterations. For the HALS
algorithm, the metric corresponds to the sum of all projected gradients of the
objective function. The algorithm also stops when it has reached 10~000
iterations if the convergence criterion is not met for the corresponding update
algorithm. This algorithm will be referred to as the Non-Negative Matrix
Factorization - Partial Endmember-Abundance Knowledge (NMF-PEAK).

The NMF-PEAK has some resemblances with other NMF-based algorithms, such as the
NMF-PPK developed by Tong \textit{et al.}
\cite{tong_nonnegative-matrix-factorization-based_2016} and the MCG-NMF
developed by Wang \textit{et al.} \cite{wang_multiple_2020}. However, the
NMF-PPK only considers a partial prior on the endmembers in its objective
function, and the scalar that weights the partial prior knowledge about the
endmembers is the same for every endmember while it can have a different value
in our approach. Moreover, the sum-to-one constraint for the abundance matrix is
not forced in our approach compared to the NMF-PPK, and the initialization is
performed by a VCA algorithm instead of the NNDSVDA method applied here. We also
propose two update algorithms for this objective function.

The MCG-NMF on the other hand considers a prior on both the endmembers and the
abundances, however the application of the algorithm is very different from our
case, for the MCG-NMF gets self-supervised information about endmembers and
abundances by using multiple clustering in the image to get known information
about the endmembers and the abundances.

\section{Validation on Simulated and Real Data}
\label{sect:validation}

\begin{figure*}[!t]
\centering
\subfloat[]{\includegraphics[width=3in]{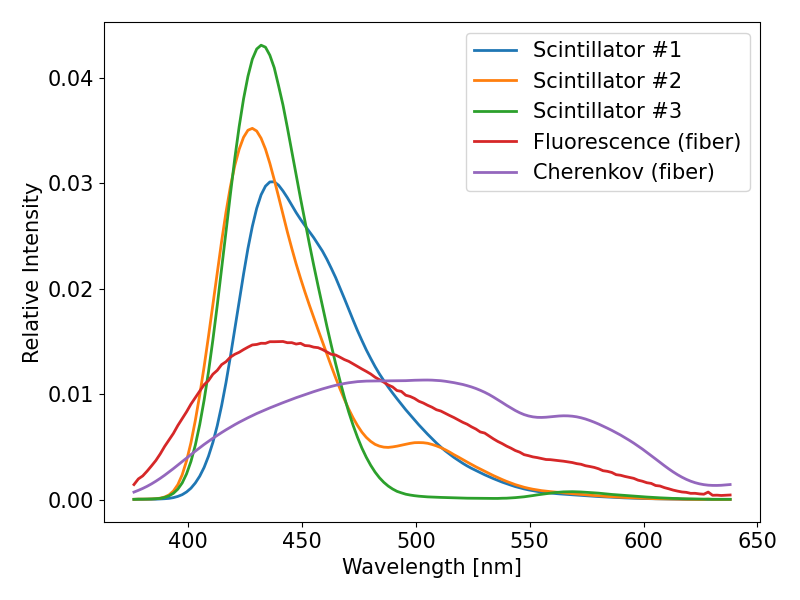}%
\label{fig:endmembers}}
\hfil
\subfloat[]{\includegraphics[width=3in]{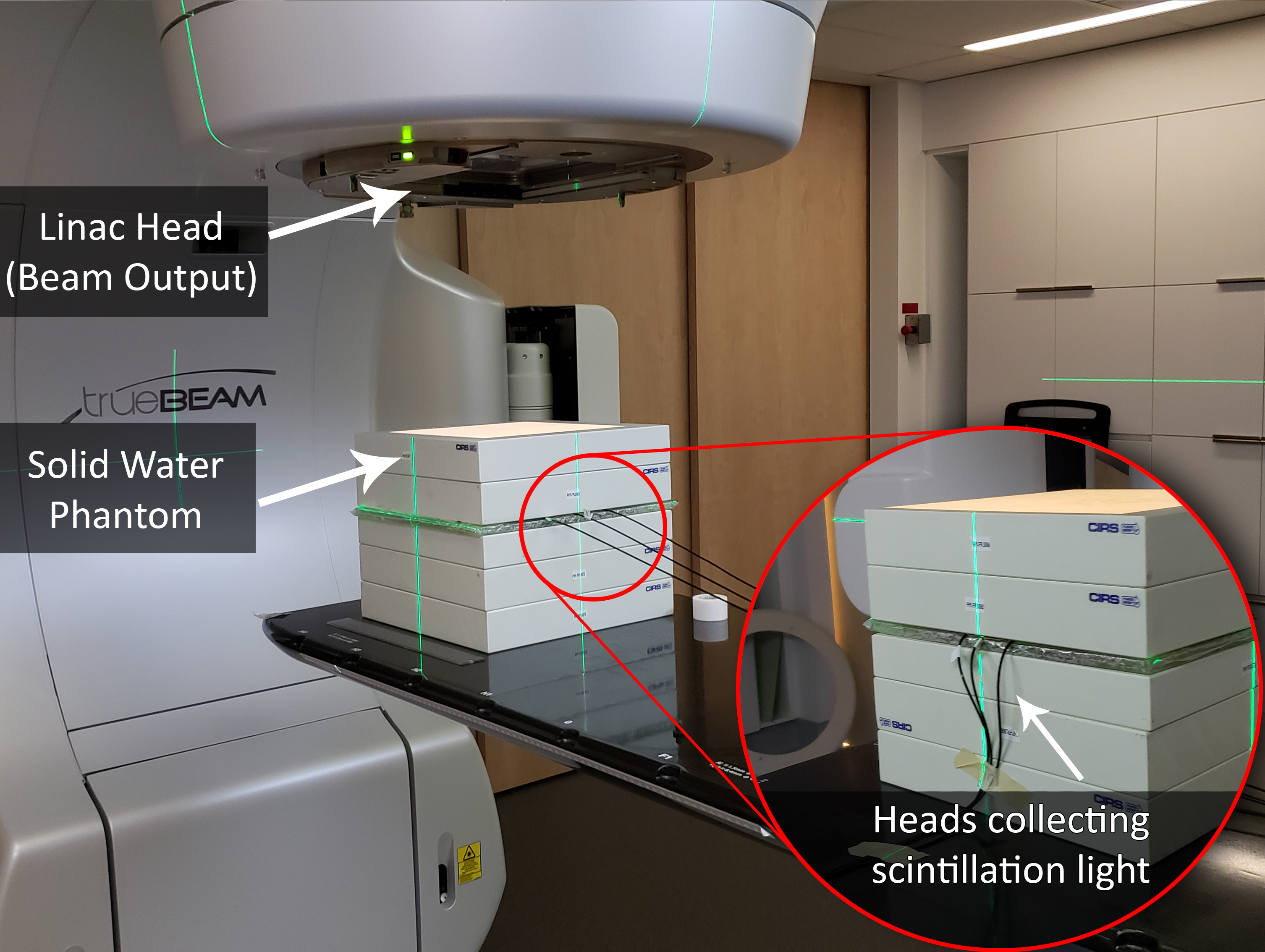}%
\label{fig:setup}}
\caption{(a) Endmembers that compose the signal measured experimentally by the mPSD. The spectra are normalized by their total intensity. The two broad spectra (endmembers 4 and 5) represent the stem spectra (Cherenkov radiation and fluorescence). (b) Experimental setup at the linac. The \textcolor{black}{probe heads with their} \textcolor{black}{scintillator} are in the water phantom and \textcolor{black}{are ultimately connected to a single optical fiber}.}
\label{fig:endmembers_mPSD}
\end{figure*}

The mPSD used for the validation of the NMF-PEAK is a 3-point scintillation
dosimeter \textcolor{black}{having 3 detachable heads connected to a single optical fiber, each head having its unique scintillator at the tip} (i.e. a dosimeter containing three scintillator elements each emitting
light with a different spectrum). The total signal measured is then composed of
5 endmembers: one distinct endmember for each scintillator element and two
endmembers produced within the stem of the detector: Cherenkov and fluorescence
as outlined in \cite{okeeffe_review_2008, therriault-proulx_nature_2013}.
All endmembers are shown in Fig.~\ref{fig:endmembers}. The experimental setup is
shown in Fig. \ref{fig:setup}. The mPSD containing the three scintillator
elements is placed inside a solid water phantom for measurements. The \textcolor{black}{three heads are}
connected to an optical fiber to collect the light produced within the mPSD to
an acquisition system located outside the treatment room. The scintillators and
fiber diameters are 1~mm, and the system used for the acquisition is the
Hyperscint HS-RP200 (Medscint inc., Quebec, Canada) \cite{jean_comparative_2021}
\cite{noauthor_home_nodate}. 

\subsection{Validation and Optimization of the Algorithm using Simulated Data}
\label{sect:validation_simul}

Simulations of the experimental calibration method are performed in order to
(1)~assess the general performance of NMF-PEAK, (2)~\textcolor{black}{identify} the best update
rule between MUR and HALS and (3)~evaluate the role of the
$\alpha_k$ and $\beta_m$ trust coefficients. The 5 endmembers measured
experimentally and shown in Fig.~\ref{fig:endmembers} are used for the
simulations and 18 synthetic factory and user calibration data with different
abundances are generated using these endmembers.

Comparisons between an estimated \textcolor{black}{endmember} $\mat{r}_{k}$ and its ground truth
\textcolor{black}{value} $\mat{r}_{k,0}$ is done using spectral angle distances (SAD) \cite{tong_nonnegative-matrix-factorization-based_2016}, defined as:

\begin{equation}
    \textnormal{SAD}_k = \arccos\left(\frac{\mat{r}_k \cdot \mat{r}_{k,0}}{||\mat{r}_k||\; ||\mat{r}_{k,0}||}\right).
\end{equation}

To evaluate the average performance of the algorithm, the mean SAD over all
\textcolor{black}{endmembers} is computed:

\begin{equation}
    \overline{\textnormal{SAD}} = \frac{1}{K}\sum_{k=1}^{K} \textnormal{SAD}_k.
\end{equation}

The metric used for the comparison of the expected ($\mat{x}_{k,0}$) and
estimated ($\mat{x}_{k}$) abundance arrays, for a given endmember $k$, is the
root-mean-square error (RMSE), defined as

\begin{equation}
    \textcolor{black}{\textnormal{RMSE}_k = \sqrt{\frac{1}{M} ||\mat{x_k} - \mat{x_{k,0}}||^2}\,.}
\end{equation}

As for the SAD, the average quantity $\overline{\textnormal{RMSE}}$ is also 
computed to evaluate the overall performance. The metric used to verify the 
accuracy of dose measurements performed by the mPSD is the mean percent error, 
defined as

\begin{equation}
    \delta D_k = \frac{1}{M} \sum_{m=1}^{M} \left(\frac{D_{k,m}}{D_{\textnormal{ref}}} - 1\right) \cdot 100\,,
\end{equation}
where $D_{k,m}$ is the dose measured for endmember $k$, for measurement $m$, which
is proportional to the abundance $x_{k,m}$, and $D_{\textnormal{ref}}$ is the reference dose
measurement for a given irradiation geometry.

\subsubsection{Comparison Between Update Rules}

\begin{figure}[!t]
\centering
\includegraphics[width=3.25in]{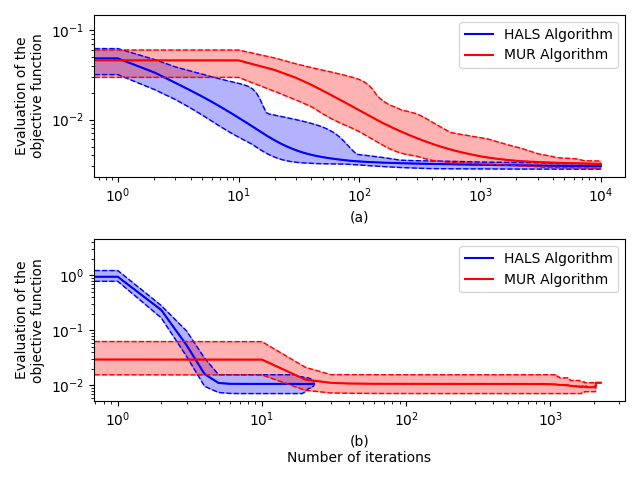}
\caption{Mean values of the objective function of the NMF-PEAK, for 100 different random datasets, as a function of the number of iterations for the MUR and HALS algorithms. The dashed lines represent the minimum and maximum values reached by the objective function for the 100 different datasets. In (a), $\mat{A}^2 = \mat{B}^2 = \mat{0}$ (which is equivalent to the basic NMF algorithm), while in (b), $\alpha_k^2 = 0$ for every endmember except for the least present endmember in the dataset provided to the algorithm, for which $\alpha_k^2 = 1$, and $\beta^2 = 1$ for every measurement.}
\label{fig:error_iter_URs}
\end{figure}

Both update rules were studied using simulated data for the NMF-PEAK and the
standard NMF algorithms (i.e. with $\mat{A}^2 = \mat{B}^2 = \mat{0}$). 
In \textcolor{black}{this simulation}, the ground truth endmembers correspond to the endmembers measured experimentally and shown in figure \ref{fig:endmembers}. The only prior on endmembers is a prior on the fluorescence spectrum, and corresponds to a spectrum acquired experimentally and having a SAD of 0.0755 with the ground truth fluorescence spectrum.
The two fluorescence spectra were not measured with the same device, which explains the non-zero SAD between the two spectra.
The use of a prior only on the fluorescence spectrum will be explained later.

To generate the simulated ground truth abundances and corresponding spectra, 18 abundance values are
sampled randomly
from a uniform distribution between 0 to 0.90 for 4 of the 5 endmembers. The
abundance of the fifth endmember, which corresponds to the fluorescence spectrum, is sampled from a uniform distribution between
0 to 0.05 (referred to as the least present endmember). Furthermore, for each
endmember, at least one simulated abundance value in the dataset is forced to
possess the maximum value. The abundances are also chosen so that the
sum of all abundances for a given measurement is 1.

The contribution of the fluorescence spectrum relative to the total signal was chosen to be very small to replicate what is observed experimentally, as shown by Therriault-Proulx \textit{et al.} \cite{therriault-proulx_nature_2013}. The small proportion of fluorescence typically makes it difficult to retrieve accurately.

To generate the prior abundances, gaussian noise with a standard deviation of 0.02 is added to the ground truth abundance matrix, resulting in a prior abundance matrix having a mean RMSE of 0.0810 with the prior abundance matrix.

The 18 mixed spectra generated for the simulation correspond to the product of the ground truth endmembers and the ground truth abundances, multiplied by a scaling factor so that their absolute intensity \textcolor{black}{corresponds} to what can be observed experimentally.
Poisson noise is also added to these simulated spectra, to simulate the presence of the shot noise that is present experimentally.
The standard deviation of this noise corresponds to the square root of the spectral signal.
These simulated spectra represent the \textit{User Calibration Data}.

Given the simulated mixed spectra and the prior on both the abundances and the fluorescence endmember, the NMF-PEAK algorithm is applied first with the MUR algorithm, and then with the HALS algorithm.
For the NMF-PEAK, we choose $\alpha_{fluo}^2 = 1$, $\alpha_{k \neq fluo} = 0$ and $\mat{B}^2 =
\beta^2\mat{I}_{[18\times 18]}$ with $\beta = 1$.
The standard NMF algorithm is also applied for comparison purposes.
The value of the objective function, at each iteration and for both \textcolor{black}{algorithms}, is shown in figure \ref{fig:error_iter_URs}.

This simulation is repeated 100 times, each time with a different \textit{User Calibration Dataset}. The mean curve is shown in figure \ref{fig:error_iter_URs}, with the upper and lower bounds represented by dashed lines.
This means that the 100 curves lie within the blue- and red-shaded regions of figure~\ref{fig:error_iter_URs}.

It can be seen from Fig. \ref{fig:error_iter_URs} (a)~that, in the case of the
basic NMF algorithm (i.e. when $\mat{A}^2 = \mat{B}^2 = \mat{0}$), the algorithm reaches the maximum number of iterations without reaching the stopping criterion for both the
HALS and the MUR algorithm. However, the HALS algorithm generally has a faster
descent than the MUR algorithm. In the case of the NMF-PEAK algorithm, it can be
seen again from figure \ref{fig:error_iter_URs} (b)~that the HALS algorithm
converges faster than the MUR algorithm. The HALS algorithm requires between 15
and 23 iterations to meet its stopping criterion, while the MUR algorithm
requires between 720 and 2240 iterations to meet its stopping criterion.
Nevertheless, both algorithms lead to objective function values between 0.00708
and 0.0156. These results are in agreement with the literature, stating that the
multiplicative update algorithms are slow to converge compared to other
algorithms such as alternating least-squares algorithms, which can converge very
rapidly depending on their implementation \cite{berry_algorithms_2007}. It is
therefore not surprising to see that the HALS algorithm converges faster than
the MUR for the NMF-PEAK. Consequently, the HALS algorithm will be used for the
remainder of this work.

\subsubsection{Effect of the prior on abundances}

The next simulation aims at studying the effect of matrix $\mat{B}$ that weights
the priors on abundances in eq.~(\ref{eq:costfct}). 
The ground truth and prior endmembers are the same as for the previous simulation, and the abundances and mixed spectra are generated the same way as well. 
However, instead of adding a gaussian noise with a constant standard deviation of 0.02 to the ground truth abundance matrix to create the prior abundance matrix, multiple prior abundance matrices are generated, each with a different gaussian noise level ranging from 0 to 18.2.
The RMSE between the ground truth and prior abundance matrices therefore ranges between 0 and 0.52.
The NMF-PEAK algorithm is applied with $\mat{A}^2 = \mat{0}$ and $\mat{B}^2 = \beta^2\mat{I}_{[18\times 18]}$, for values of $\beta^2$ ranging from 0 to 100.
Again here, this whole process is repeated 100 times with a different \textit{User Calibration Dataset} each time.

\begin{figure}[!t]
\centering
\includegraphics[width=3.25in]{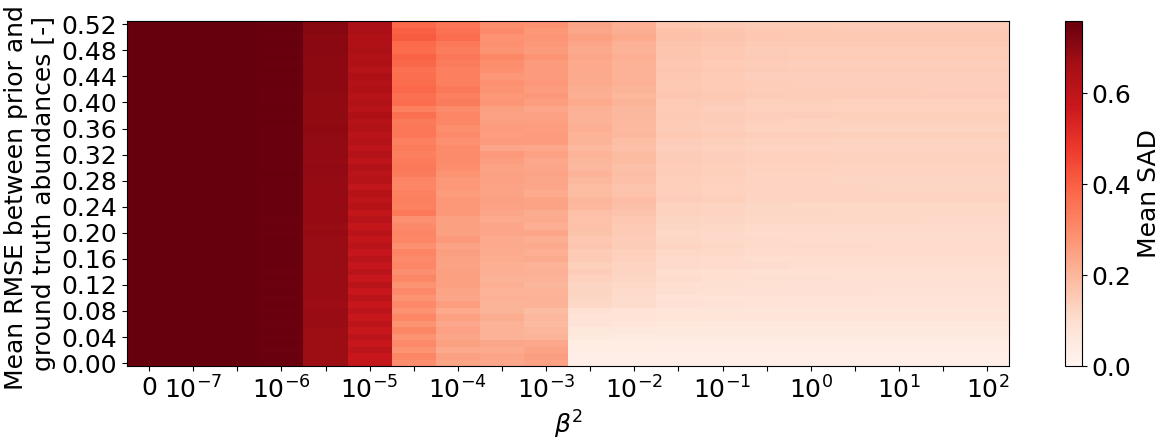}
\caption{Mean SAD between the expected and estimated spectra using the NMF-PEAK for different values of $\beta^2$ and different RMSE between the prior abundances and the ground truth abundances.}
\label{fig:simul_beta}
\end{figure}

The mean SAD between the retrieved and the ground truth endmembers is shown in figure \ref{fig:simul_beta}, as a function of the mean RMSE between the ground truth and prior abundances and as a function of $\beta^2$.
For $\beta^2 = 0$, we have the standard NMF algorithm, which leads to the
greatest SAD without surprise. As $\beta^2$ increases, the SAD generally
diminishes slowly, then there is a sharp drop of the SAD for $\beta^2 \approx
5.5\times 10^{-5}$, followed by a small decrease of the SAD with increasing
$\beta^2$. The mean SAD increases when the RMSE increases because the prior is
farther from the real abundances. However, it is interesting to note that, even
for high deviations, there is not a significant increase of the SAD with
$\beta^2$, so even if we force an abundance matrix that is not highly accurate,
it still guides the convergence towards a better solution than if no prior was
used. This result demonstrates the necessity to use a prior on the abundances
for the calibration of mPSDs. With a prior only on the abundances, the results
can be really improved compared to the ones that would be obtained with the
standard NMF, even if the prior has a mean RMSE around 0.52 with the ground
truth abundances.

\subsubsection{Effect of the prior on endmembers}

When the prior abundances are well estimated, the error on the endmembers retrieved by the NMF-PEAK algorithm is small.
However, when an endmember is not
present in big proportions in the dataset, as in the case of the fluorescence having no more than 5\% of relative abundance, the algorithm can be
helped by providing prior information on this endmember.

\begin{figure}[!t]
\centering
\subfloat[]{\includegraphics[width=3in]{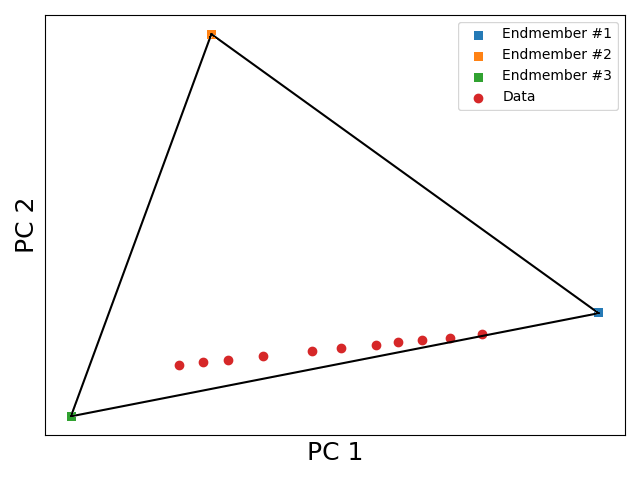}\label{fig:pca}}\\
\subfloat[]{\includegraphics[width=3.25in]{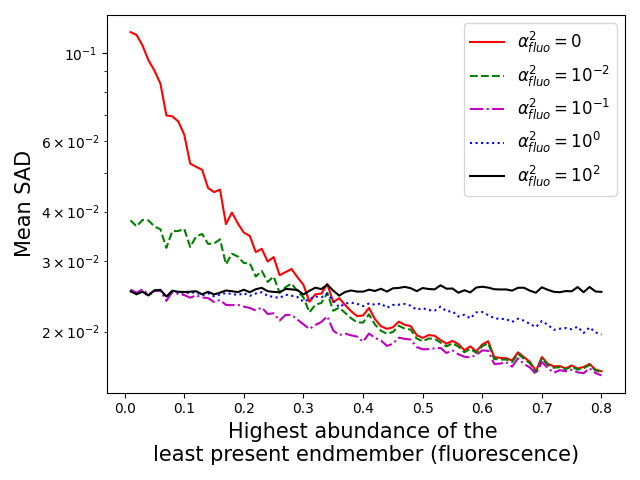}\label{fig:SAD_fct_p}}
\caption{(a) Example of a PCA performed on a 3-endmembers dataset where one of the endmembers has very low abundances. (b) Mean SAD between the ground truth and estimated spectra as a function of the highest abundance of the least present endmember in the calibration dataset.}
\label{fig:pca_SADfctP}
\end{figure}

To better visualize the problem, let's consider a 3-endmembers unmixing problem.
Fig. \ref{fig:pca} shows an example of a calibration dataset for which one of
the endmembers has very low abundances.
To produce this figure, a principal component
analysis (PCA) \cite{jolliffe_principal_2016} was performed on the dataset and
the first two principal components are plotted. The endmembers form a simplex
inside of which the mixed spectra lie. According to this figure, it is understood
that all the calibration data are principally composed of endmembers 1 and 3,
and the abundance of endmember 2 is very small. In such a case, the use of a
prior knowledge on the least abundant endmember can help retrieving this
endmember, as well as the others. 

Going back to the original 5-endmembers unmixing problem, Fig. \ref{fig:SAD_fct_p} shows the mean SAD between the ground truth and
estimated endmembers as a function of the highest abundance of the least present
endmember in the calibration data and for different values of $\alpha_{fluo}^2$.
The fluorescence spectrum, which is the least present endmember, has relative
abundances between 0 and 0.01 for every \textcolor{black}{measurement} of the dataset, except for
one measurement, where the maximum is imposed to a value ranging from 0.01 to
0.90. For this particular measurement, the relative abundances of the other
endmembers therefore decrease as the abundance of the least present endmember
increases.

The ground truth and prior endmembers are the same as for the first simulation, and the abundances and mixed spectra are generated the same way as well.
The NMF-PEAK algorithm is applied with $\alpha_{k \neq fluo} = 0$ and $\mat{B}^2 =
\beta^2\mat{I}_{[18\times 18]}$ with $\beta = 1$.
The simulation is repeated 100 times, and the average
curves are presented in figure \ref{fig:SAD_fct_p} for each given value of $\alpha_{fluo}^2$.

From Fig. \ref{fig:SAD_fct_p}, it can be seen that the mean SAD tends to increase
when all the abundances of one endmember become very small, which is usually
experimentally the case for the fluorescence spectrum. To address this problem, the use of a prior on this spectrum ($\alpha_k^2
\neq 0$) is essential to help accurately retrieving this spectrum. However, when the least present
endmember has at least one measurement with a relatively high abundance and a
high value for $\alpha_k^2$, the use of the prior on this particular endmember
could lead to a greater SAD, because the prior endmember provided to the
algorithm is not actually the real endmember.
\textcolor{black}{It is also possible to see on figure \ref{fig:SAD_fct_p} that when the value for $\alpha_k^2$ is really large, the algorithm becomes too restrictive and the retrieved endmember corresponds most likely to the prior, regardless of the abundance of this endmember in the dataset provided to the algorithm. That is why the variation of $\alpha_k^2$ is so little in that situation. If the prior endmember does not correspond perfectly to the actual endmember, then large values for $\alpha_k^2$ won’t lead to the actual endmember.}
This figure also shows that
$\alpha_k^2 \geq 10^{-1}$ is necessary to minimize the mean SAD if the highest
abundance of the least present endmember is very low, but that a smaller value
for $\alpha_k^2$ can be used if the spectrum is present in enough great
proportions. For this particular case, with $\alpha_{fluo}^2 = 10^{-1}$, we
obtain the smallest possible SAD for a given value for the highest abundance of
the \textcolor{black}{least} present endmember. These results demonstrate the necessity to include a
prior on the spectra for which the abundance is really small in the calibration
data.

\subsection{Experimental \textcolor{black}{Determination} of the \textcolor{black}{Priors}}

\begin{figure*}[!t]
\centering
\subfloat{\includegraphics[width=6in]{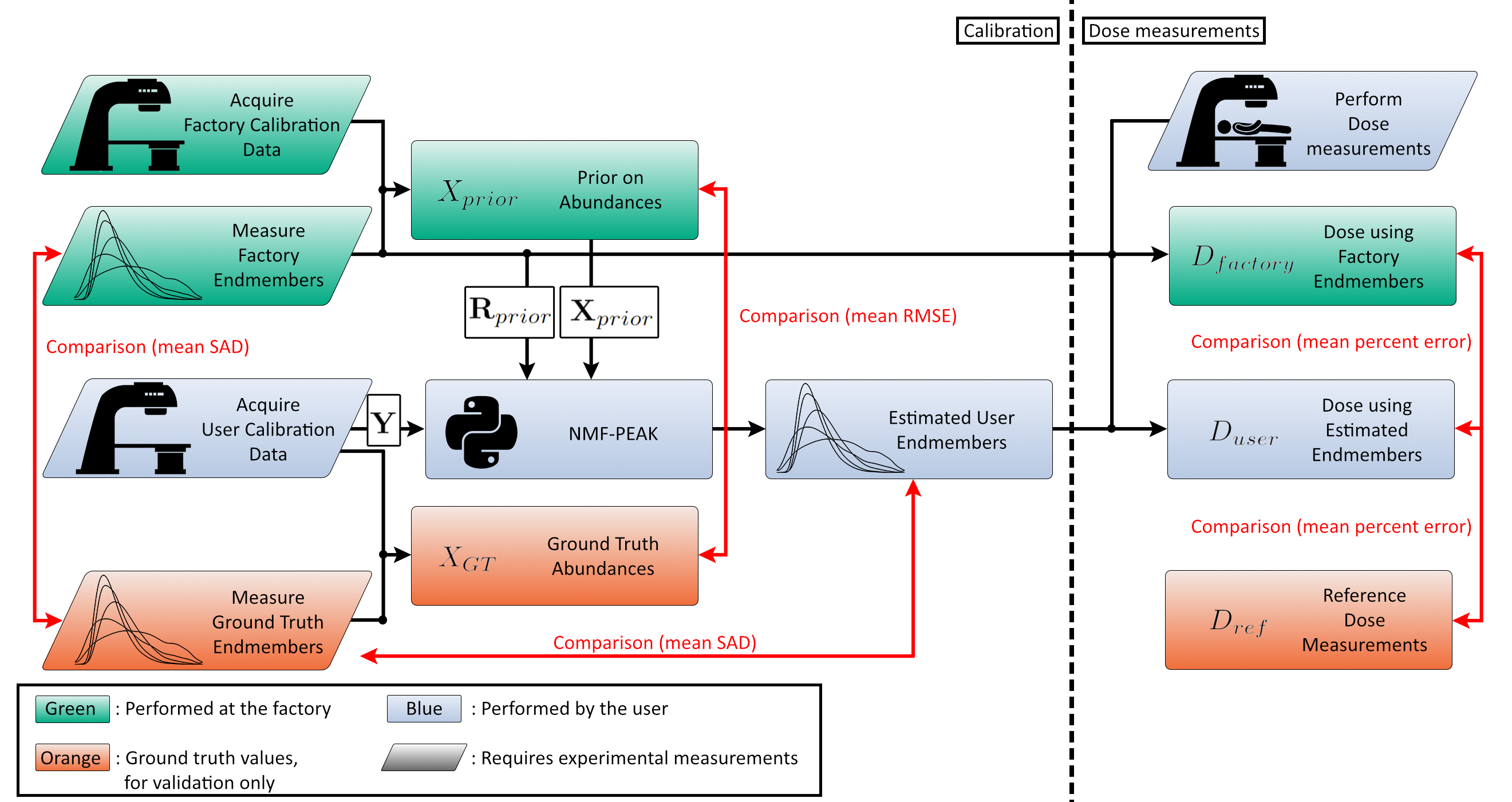}%
}
\caption{Experimental process applied in this study.}
\label{fig:process}
\end{figure*}

Before being assembled to form a mPSD, each \textcolor{black}{detachable head of the mPSD} was
\textcolor{black}{successively} coupled to \textcolor{black}{a single-head} optical fiber to form \textcolor{black}{a} \textcolor{black}{probe} with only a single
scintillator element. Each of these single scintillator probes only contains three
endmembers: the scintillator spectrum, the Cherenkov \textcolor{black}{spectrum} and the
fluorescence \textcolor{black}{spectrum}.
The endmembers of each single scintillator probe \textcolor{black}{were} determined \textcolor{black}{following the procedure described in appendix \ref{app:long_calib}}.
These measurements are referred to as the \textit{Factory Endmembers} in figure
\ref{fig:process}. Then the simple calibration routine \textcolor{black}{that will have to be applied by the end user} was applied to \textcolor{black}{the} single
scintillator probes in order to get the \textit{Factory Calibration Data}. The
measurements performed with these 3 independent probes are therefore used to
determine the prior endmembers and prior abundances \textcolor{black}{for the end user}.

\subsection{Validation on Experimental Data}

Now that we have demonstrated the performances of our approach and the influence
of matrices $\mat{A}$ and $\mat{B}$ on simulated data, we apply the NMF-PEAK on
experimental data. As described previously, the priors on endmembers and
abundances come from measurements performed with individual probes, while the
NMF-PEAK calibration is applied on a single mPSD having 3 scintillators. 

To perform the simple calibration
routine \textcolor{black}{(the first step of the two-steps process performed by the user)}, the mPSD is placed at 10~cm depth in a solid water phantom, \textcolor{black}{at the isocenter of the linac,} and is
irradiated under a 6~MV photon beam. \textcolor{black}{A set of 18 measurements composed of irradiations with square fields of $2\times 2$~cm$^2$, $5\times 5$~cm$^2$, $8\times 8$~cm$^2$, $10\times 10$~cm$^2$, $20\times 20$~cm$^2$ and $30\times 30$~cm$^2$ are made for each scintillator separately in order to vary the proportion of
light emitted by each component.} These measurements are referred to as the
\textit{User Calibration Data} in figure \ref{fig:process}.

\textcolor{black}{The user then applies the NMF-PEAK algorithm (the second step of the two-steps process performed by the user).} 
The \textit{User Calibration Data} acquired by \textcolor{black}{performing} the simple calibration routine
with the mPSD are fed to the NMF-PEAK algorithm, along with the priors on endmembers
and abundances obtained from the \textit{Factory Calibration Data} and the
\textit{Factory Endmembers}, leading to the \textit{Estimated User Endmembers}
as seen in figure \ref{fig:process}.

\textcolor{black}{The actual endmembers of the mPSD were also measured following the procedure described in appendix \ref{app:long_calib}.}
These measurements are referred to as the \textit{Ground Truth
Endmembers} in figure \ref{fig:process}. Performing these ground truth
measurements \textcolor{black}{can be time consuming (up to 1h)} and is not always possible because some linac\textcolor{black}{s}
are not equipped with an X-ray tube.

\textcolor{black}{To verify the accuracy of the calibration
using either the \textit{Estimated User Endmembers}, the \textit{Ground Truth
Endmembers} or the \textit{Factory Endmembers} directly, partial
or total irradiation of all or some of the scintillators of the mPSD is
performed in order to have independent dose measurements that are compared to reference clinical dose values which are precisely known.}
\textcolor{black}{A total of 22 measurements were performed for each scintillator, under a 6~MV photon beam at the isocenter of the linac, with square fields varying from $2\times 2$~cm$^2$ to $30\times 30$~cm$^2$. Half of these measurements were performed with only one head at a time in the field (i.e. with only one scintillator), and the other half was performed with the two other heads at $\pm 2.5$~cm from each other (i.e. with all scintillators partially or totally in the field).}

Table \ref{tab:exp_results} shows the SAD, for each of the 5 endmembers, between
the ground truth endmembers and the factory endmembers, the estimated endmembers
using the basic NMF and the estimated endmembers using the NMF-PEAK. The mean
percent error on dose measurements, compared to the reference dose measurements,
is also obtained using the factory endmembers, the estimated endmembers using
the basic NMF and the estimated endmembers using the NMF-PEAK algorithm. Here,
the basic NMF algorithm (i.e. with $\mat{A} = \mat{B} = \mat{0}$) and the NMF-PEAK algorithm (with non-zero matrices for $\mat{A}$ and $\mat{B}$) are applied using 18
calibration data. The mean RMSE between the ground truth abundance matrix and
the prior abundance matrix is 0.0276 in that case. The calibration is also
performed without applying any algorithm, just by using directly the factory
endmembers. For the NMF-PEAK algorithm, the matrices $\mat{A}$ and $\mat{B}$
were optimized in order to get the best results on the estimated spectra.

\begin{table}[!t]
\renewcommand{\arraystretch}{1.3}
\caption{SAD between endmembers and error on dose measurements. The mean RMSE \textcolor{black}{between ground} truth and prior abundances is 0.0276}
\label{tab:exp_results}
\centering
\resizebox{.45\textwidth}{!}{%
\begin{tabular}{c|ccclccc}
\hline\hline
                     & \multicolumn{3}{c}{\textbf{\begin{tabular}[c]{@{}c@{}}SAD relative to\\ Ground Truth Endmembers\end{tabular}}} &  & \multicolumn{3}{c}{\textbf{\begin{tabular}[c]{@{}c@{}}Dose Mean Percent Error\\ relative to Reference Dose Measurements [\%]\end{tabular}}} \\
Endmember            & Factory                             & NMF                               & NMF-PEAK                             &  & Factory                   & NMF                    & NMF-PEAK                             \\ \cline{1-4} \cline{6-8} 
Scintillation \#1    & 0.0718                               & 0.3140                             & 0.0032                                &  & $(3 \pm 11)$     & $(-7 \pm 32)$    & $(-0.17 \pm 0.28)$                                \\
Scintillation \#2    & 0.0252                               & 1.1822                             & 0.0056                                &  & $(-8 \pm 15)$    & $(7 \pm 76)$     & $(0.70 \pm 0.69)$                                \\
Scintillation \#3    & 0.1650                               & 0.8856                             & 0.0056                                &  & $(2 \pm 3)$      & $(-4 \pm 79)$    & $(0.24 \pm 0.81)$                                \\
Fluorescence (fiber) & 0.0944                               & 0.7335                             & 0.0944                                &  & -                      & -                      & -                                \\
Cherenkov (fiber)    & 0.0265                               & 0.8061                             & 0.0038                                &  & -                      & -                      & -                                \\ \cline{1-4} \cline{6-8} 
\textbf{Mean}        & 0.0766                               & 0.7843                             & 0.0225                                &  & $(-1 \pm 11)$    & $(-2 \pm 66)$    & $(0.25 \pm 0.73)$                                \\ \hline\hline
\end{tabular}%
}
\end{table}

From this table, it is possible to see that the SAD between the factory and the
ground truth endmembers has a significant effect on dose measurements,
justifying the need to perform a new calibration even if the endmembers were
previously measured in different conditions. 
\textcolor{black}{This SAD between the factory and the ground truth endmembers can be explained by the sensitivity of the optical system. First, the 3 detachable heads of the mPSD are successively connected to a different single-head optical fiber during the acquisition of the priors, like if they were 3 independent probes. The effect of this connection, as well as the connection between the collecting fiber and the optical reader, induces small changes in the shape of the endmember spectra. The spectral attenuation of the collecting fiber can also have an effect since the fiber between the detachable heads and the optical reader is not the same in both cases. Finally, the prior fluorescence and Cherenkov spectra correspond to the average spectra measured for the 3 independent heads.}

It can be seen \textcolor{black}{that} if the factory
spectra were directly used instead of performing a user calibration, the error
could be greater than 10\%. Applying the basic NMF algorithm causes even larger
errors than using the factory calibration. Of all the \textcolor{black}{scenarios} tested, using
NMF-PEAK algorithm combined with the simple calibration routine provided the
best results. Moreover, spectra estimated with NMF-PEAK lead \textcolor{black}{to} average dose
\textcolor{black}{measurement} errors of $(0.25 \pm 0.73)$\%. It is important to mention here that
if the NMF-PEAK algorithm had been applied without optimizing matrices
$\mat{A}^2$ and $\mat{B}^2$, and by taking instead $\alpha_{fluo}^2 = 10^{-1}$,
$\alpha_{k\neq fluo}^2 = 0$ and $\mat{B}^2 = \beta \mat{I}_{[18\times 18]}$,
with $\beta = 1$, the average error on dose measurements would have been $(0.45
\pm 1.82)$\%, with a mean SAD of 0.0265 relative to the ground truth endmembers,
which is still a lot better than if no calibration was performed. These
parameters were chosen accordingly to the simulations previously done. \textcolor{black}{These
experimental results justify} the use of the NMF-PEAK algorithm for the
calibration of mPSDs.

\section{Conclusion}
\label{sect:conclusion}

The use of \textcolor{black}{an} NMF-based algorithm in the field of medical physics, for the calibration of mPSDs, allows to perform an easy and precise calibration process for mPSDs. The new NMF-based algorithm developed in this study, the NMF-PEAK, allows to include partial prior knowledge on both the endmembers and the abundances. The abundances can be estimated from a reference calibration routine and provided to the algorithm to help it converge to a very precise solution. If needed, some endmembers can also be provided as a prior, with a different degree of certitude for each endmember to retrieve. This algorithm has the advantage to be easy to implement from an already existing NMF algorithm and allows to give a different degree of uncertainty on individual endmembers and abundance arrays in the prior array. For future work, the use of machine learning algorithms could be investigated for the calibration of mPSDs, without the use of a particular prior for each calibration routine.


%

\appendices
\section{Proof of the Non-Increasing Property of the Multiplicative Update Rule Algorithm}
\label{app:proof_nip_MUR}

As mentioned in section \ref{sect:approach}, the update rules of the multiplicative update rule (MUR) algorithm are given by

\begin{align}
    \mathbf{R}_{lk}^{(n+1)} &= \mathbf{R}_{lk}^{(n)} \cdot \left(\frac{(\mat{YX}^T)_{lk}^{(n)} + \alpha_k^2(\mat{R}_\mathrm{prior})_{lk}}{(\mat{RXX}^T)_{lk}^{(n)} + \alpha_k^2\mat{R}_{lk}^{(n)}}\right)\,,
    \label{eq:MUR_R_02}
    \\
    \mathbf{X}_{km}^{(n+1)} &= \mathbf{X}_{km}^{(n)} \cdot \left(\frac{(\mat{R}^T\mat{Y})_{km}^{(n)} + \beta_m^2(\mat{X}_\mathrm{prior})_{km}}{(\mat{R}^T\mat{RX})_{km}^{(n)} + \beta_m^2\mat{X}_{km}^{(n)}}\right)\,.
    \label{eq:MUR_X_02}
\end{align}

To prove the non-increasing property of the algorithm, the method \textcolor{black}{developed} by Lee and
Seung~\cite{lee_algorithms_nodate} was used in order to get the following
non-increasing criteria for both update rules, for given vectors $\mat{v_R}_{[LK \times 1]}$ and $\mat{v_X}_{[KM \times 1]}$:

\begin{align}
    \mat{v_R}^T\left(\mat{D_R} - \mathcal{H}_\mat{R}[F(\mat{R},\mat{X})]\right)\mat{v_R} &= \mat{v_R}^T\mat{S_R}\mat{v_R} \geq 0\,, \label{eq:pos_semi_def_R} \\
    \mat{v_X}^T\left(\mat{D_X} - \mathcal{H}_\mat{X}[F(\mat{R},\mat{X})]\right)\mat{v_X} &= \mat{v_X}^T\mat{S_X}\mat{v_X} \geq 0\,,
\label{eq:pos_semi_def_X}
\end{align}

\noindent
where $\mat{D_R}$ and $\mat{D_X}$ are diagonal matrices having entries $d_{lk} = 1/\gamma_{lk}$ and $d_{km} = 1/\gamma_{km}$ respectively. 
In equations~(\ref{eq:pos_semi_def_R}) and (\ref{eq:pos_semi_def_X}), $\mathcal{H}_\mat{R}[F(\mat{R},\mat{X})]$ (dimensions $LK \times LK$) and $\mathcal{H}_\mat{X}[F(\mat{R},\mat{X})]$ (dimensions $KM \times KM$) are the Hessian matrices of $F(\mat{R},\mat{X})$, for a constant $\mat{X}$ and for a constant $\mat{R}$ respectively:

\begin{align}
    \mathcal{H}_\mat{R}[F(\mat{R},\mat{X})] &= 
    \left(
    \begin{array}{ccc}
    \mathcal{R} & \cdots & \mathbf{0}_{[K \times K]} \\
    \vdots & \ddots & \vdots \\
    \mathbf{0}_{[K \times K]} & \cdots & \mathcal{R}
    \end{array}
    \right)\,,
    \label{eq:Hessian_R} \\
    \mathcal{H}_\mat{X}[F(\mat{R},\mat{X})] &=
    \left(
    \begin{array}{ccc}
    \mathcal{X}_1 & \cdots & \mathbf{0}_{[K \times K]} \\
    \vdots & \ddots & \vdots \\
    \mathbf{0}_{[K \times K]} & \cdots & \mathcal{X}_M
    \end{array}
    \right)\,.
    \label{eq:Hessian_X}
\end{align}

\noindent
In equations~(\ref{eq:Hessian_R}) and (\ref{eq:Hessian_X}), the sub-matrices on
the diagonal are given by:

\begin{align}
    \mathcal{R}_{[K \times K]} &= \mat{XX}^T + \mat{A}^2 \\
    \mathcal{X}_{m\,[K \times K]} &= \mat{R}^T\mat{R} + \beta_m^2 \mat{I}_{K}\,,
\end{align}

\noindent
where $\mat{I}_K$ is the identity matrix of size $K \times K$. Matrices
$\mathcal{R}$ and $\mathcal{X}$ directly correspond to the Hessian matrices if
$L = 1$ and $M = 1$ respectively.

It is possible to prove that $\mat{S_R}$ and $\mat{S_X}$ are positive
semi-definite for the chosen step size by evaluating the positive
semi-definiteness \textcolor{black}{of matrices} $\mat{U_RS_RU_R}$ and $\mat{U_XS_XU_X}$, where
$\mat{U_R}_{[LK \times LK]}$ and $\mat{U_X}_{[KM \times KM]}$ are diagonal
matrices having respectively $\mat{R}_{lk}$ and $\mat{X}_{km}$ on their
diagonal. If matrices $\mat{U_RS_RU_R}$ and $\mat{U_R}$ are Hermitian diagonally
dominant matrices with real non-negative entries on their diagonal, they are
therefore positive semi-definite and it implies that $\mat{S_R}$ is also
positive semi-definite. The same logic applies for $\mat{S_X}$. By assuming that
$\mat{R}$ and $\mat{X}$ are non-negative matrices, it is easy to see that
$\mat{U_R}$ and $\mat{U_X}$ are positive semi-definite. In the case of
$\mat{U_RS_RU_R}$ and $\mat{U_XS_XU_X}$, it is easy to see that $\mat{U_RS_RU_R}
= (\mat{U_RS_RU_R})^T$ and that $\mat{U_XS_XU_X} = (\mat{U_XS_XU_X})^T$, and
that the matrices are therefore Hermitian. The entries on their diagonal is
given by:

\begin{align}
    \mat{U_RS_RU_R}:\qquad & \sum_{k' \neq k} \mat{R}_{lk}\mat{R}_{lk'}(\mat{XX}^T)_{k'k} \geq 0\,,
    \label{eq:S_R_diag} \\
    \mat{U_XS_XU_X}:\qquad & \sum_{k' \neq k} \mat{X}_{km}\mat{X}_{k'm}(\mat{R}^T\mat{R})_{k'k} \geq 0\,,
    \label{eq:S_X_diag}
\end{align}

\noindent
which is equal to the sum of the magnitudes of the non-diagonal entries for a
given row, and which makes these matrices diagonally dominant as well. These
conditions are always true because the entries of $\mat{R}$ and $\mat{X}$ must
be positive. It is interesting to note that the terms taking into account
the prior knowledge on $\mat{R}$ and $\mat{X}$ in the objective function do not
influence the non-increasing conditions. Therefore, $\mat{S_R}$ and $\mat{S_X}$
are positive semi-definite and the algorithm will not diverge for $\mat{R}$ and
$\mat{X}$.

\section{Experimental \textcolor{black}{Process} for the \textcolor{black}{Extensive Calibration} of PSDs}
\label{app:long_calib}

\textcolor{black}{As described in the paper, one way to obtain the prior on endmembers ($\mat{R}_\mathrm{prior}$) is to measure them at the factory, for example with a different reader or a different probe having the same scintillators. This is what was performed in this study, where the detachable heads of the mPSD were successively coupled to a different single-head optical fiber. The priors were then determined with 3 independent probes, each probe representing one head of the mPSD. The extensive calibration was performed in this context to measure the \textcolor{black}{3} endmembers of these 3 independent probes.}

\textcolor{black}{Here is how the extensive calibration is usually performed.} To experimentally determine \textcolor{black}{2 of the 3} endmembers, \textcolor{black}{namely the scintillation and the fluorescence components,} the X-ray tube of a Varian
TrueBeam linear accelerator (Varian Medical Systems, Palo Alto, USA), also
called a linac, is used to generate a low-energy radiation beam below the
threshold for producing Cherenkov radiation. By collimating this beam of
radiation \textcolor{black}{on the scintillator}, it is possible to measure \textcolor{black}{the scintillation} spectrum. \textcolor{black}{The fluorescence spectrum is measured the same way, by irradiating a large amount of clear optical fiber.} The
Cherenkov spectrum, on the other hand, is obtained by irradiating the dosimeter
with a 6~MV photon beam delivered by the same linac under conditions that first
maximize and then minimize its production. A subtraction of these two
measurements is made to isolate the Cherenkov spectrum from the other
endmembers. \textcolor{black}{This process allows to retrieve the 3 endmembers that are present in a single-point PSD, but can be time consuming (up to 1h) and is not always possible because some \textcolor{black}{linacs} are not equipped with an X-ray tube that can produce low-energy X-rays.}

\section*{Acknowledgment}

The authors declare the following financial interests/personal relationships which may be considered as potential competing interests with the work reported in this paper. Simon and \textcolor{black}{François} are \textcolor{black}{co-founders} of Medscint, and Boby is an employee at Medscint.

The authors would like to thank the NSERC for the funding of this project.
The authors would also like to thank the team at Medscint inc. for their help in constructing the mPSD for the study, as well as for their support during the experimental measurements. 
The authors would finally like to thank Patrick Dallaire at NQB (Quebec, Canada) for his support in the project.

\ifCLASSOPTIONcaptionsoff
  \newpage
\fi



\bibliographystyle{unsrt}
\bibliography{IEEEabrv,bibtex/blessard_all_2023-03-19}
%





\end{document}